\documentstyle[aps]{revtex}
\begin{document}
\title{Strategies and Networks for State-Dependent Quantum Cloning}
\author{Anthony Chefles\thanks{e-mail: tony@phys.strath.ac.uk} and Stephen M. Barnett}
\address{Department of Physics and Applied Physics, \\ University of Strathclyde, Glasgow G4 0NG, Scotland}
\input epsf
\epsfverbosetrue

\def\id{{\leavevmode\hbox{\small1\kern-3.2pt\normalsize1}}}%

\maketitle
\thanks{PACS: 03.65.Bz, 03.67.-a, 03.67.Hk}

\begin{abstract}
State-dependent cloning machines that have so far been considered either deterministically copy a set of states approximately, or probablistically copy them exactly.  In considering the case of two equiprobable pure states, we derive the maximum global fidelity of $N$ approximate clones given $M$ initial exact copies, where $N>M$.  We also consider strategies which interpolate between approximate and exact cloning.  A tight inequality is obtained which expresses a trade-off between the global fidelity and success probability.  This inequality is found to tend, in the limit as $N{\rightarrow}{\infty}$, to a known inequality which expresses the trade-off between error and inconclusive result probabilities for state-discrimination measurements.  Quantum-computational networks are also constructed for the kinds of cloning machine we describe.  For this purpose, we introduce two gates: the distinguishability transfer and state separation gates.  Their key properties are described and we show how they may be decomposed into basic operations.
\end{abstract}

\section{Introduction}
\renewcommand{\theequation}{1.\arabic{equation}}
\setcounter{equation}{0}

The no-cloning theorem of Wootters, Zurek\cite{WZurek} and Dieks\cite{Dieks1} shows that it is impossible to devise a means of perfectly replicating an arbitrary, unknown quantum state. While such an ideal cloning machine cannot be constructed, this important result does not prohibit cloning strategies which have a limited degree of success.  Numerous interesting contributions have been made to this subject by many authors, showing that the no-cloning theorem is not the last word on the matter.

All quantum cloning machines, except those designed to copy only orthogonal states\cite{WZurek,Dieks1,Mozyrsky}, are fallible.  Such machines can be divided into two categories: universal and state-dependent cloners.  The first kind produce approximate copies of a completely unknown quantum state.  This idea was proposed originally by Bu{\u{z}}ek and Hillery\cite{Buzek1} who showed how to construct a device which can copy any quantum state of a qubit equally well, where the figure of merit is the local fidelity.  This is the square overlap between the state of one of the approximate copies produced by the machine, and the exact state.  The Bu{\u{z}}ek-Hillery cloning machine was later shown to be optimal by Gisin and Massar\cite{Gisin} and Bru{\ss} et al\cite{Bruss1}, who demonstrated that these clones attain the maximum possible local fidelity.  More recently, the optimum universal cloning problem for multi-level quantum systems has been solved by Werner\cite{Werner} and Keyl and Werner\cite{Keyl}.

The second category is that of state-dependent cloning machines.  These are designed to reproduce only a finite number of states, and here we can identify two sub-categories.  Approximate state-dependent cloning machines were examined first, by Hillery and Bu{\u{z}}ek\cite{Hillery}.  These machines deterministically generate approximate clones of states belonging to a finite set.  Such devices have also been considered by Bru{\ss} et al\cite{Bruss2}.  More recently, it was discovered by Duan and Guo that a set of non-orthogonal, but linearly-independent pure states can sometimes be {\em exactly} cloned\cite{Duanguo1,Duanguo2}.  For non-orthogonal states, there is a non-zero probability that the cloning attempt fails.  However, it is always possible to tell if it has been a success and under some circumstances, this latter approach may be more useful.

Several results have been obtained for optimum cloning of two states.  For approximate cloning, Bru{\ss} et al \cite{Bruss2} have determined the maximum local and global fidelities for making two copies of an unknown state belonging to a known pair having equal a priori probabilities.  The global fidelity is the average square-overlap between the complete state of the system of approximate clones and that of the exact clones.  Duan and Guo\cite{Duanguo1} have found the maximum success probability if these copies are required to be exact.  More recently, we have found the maximum success probability if we have $M$ initial copies of the state, and wish to obtain $N$ copies, where $N>M$, and have shown how this result is related to unambiguous state identification\cite{Me1}.

In this paper, we obtain more general results relating to approximate cloning, and investigate properties of cloning machines which are hybrids of both types of state-dependent cloning.  In section II, we show that the use of the global fidelity as the figure of merit for approximate cloning establishes an important link between cloning and state identification, at least if the states to be copied are linearly-independent.   We then solve the complete optimisation problem for two states.  We find the maximum global fidelity for $M{\rightarrow}N$ cloning where the states have arbitrary a priori probabilities.  The maximum global fidelity obtained converges to the well-known Helstrom bound\cite{Helstrom} of quantum detection theory as $N{\rightarrow}{\infty}$.

In section III, we develop the idea of hybrid cloning machines.  Exact cloning is a probabilistic process.  We show that probablistic exact cloning and deterministic approximate cloning may be viewed as the extremal cases of a more general family of strategies, characterised by a trade-off between the success probability and the global fidelity.  This compromise relation is found, in the limit of infinite cloning, to tend to a known relationship which expresses a trade-off between the probabilities of obtaining erroneous and inconclusive results in the theory of state identification\cite{Me1,Me2}.

The remainder of the paper is concerned with issues relating to the realisation of state-dependent cloning machines.  We restrict our attention to qubits, and take the point of view that, since multiparticle cloning necessarily involves collective interactions involving many particles, it is most convenient to decompose these operations in terms of local or pairwise operations, so that they may be suitable for implementation on a quantum computer.  In section IV, we introduce two gates, the distinguishability transfer and state separation gates, and show how they can be used to build a network for exact cloning.  We use the same gates, in section V, to construct networks for optimal approximate and hybrid cloning machines.  In section VI, the distinguishability transfer and state separation gates are decomposed in terms of basic operations, that is, local unitary transformations and controlled-NOT gates.

\section{State-Dependent Approximate Cloning Machines}
\renewcommand{\theequation}{2.\arabic{equation}}
\setcounter{equation}{0}

Consider a set of $K$ non-orthogonal quantum states
$|{\psi}_{j}{\rangle}$.  If there are $M$ quantum systems, all of which are prepared in the same state, which is an unknown member of this set, then the possible states of the combined system are the $M$-fold tensor products
\begin{equation}
|{\psi}^{M}_{j}{\rangle}=|{\psi}_{j}{\rangle}_{1}{\otimes}{\ldots}{\otimes}|{\psi}_{j}{\rangle}_{M}.
\end{equation}
Our aim is to manufacture, to the best approximation, a larger
number of $N>M$ copies of the unknown state.  We introduce
a further system composed of $N-M$ subsystems prepared in
some neutral state $|{\chi}{\rangle}$ and transform the state
$|{\psi}^{M}_{j}{\rangle}|{\chi}{\rangle}$ into one approximating $N$
exact copies of the state $|{\psi}_{j}{\rangle}$, this being the $N$-fold tensor product state $|{\psi}^{N}_{j}{\rangle}=|{\psi}_{j}{\rangle}_{1}{\otimes}{\ldots}{\otimes}|{\psi}_{j}{\rangle}_{N}$.

We require the cloning operation to be deterministic, that is, the clones, though approximate, are generated on demand.  Deterministic exact cloning is impossible, so we require a figure of merit which characterises how closely our copies resemble exact copies.  One interesting measure of the quality of the final states, which we denote by $|{\Phi}_{j}{\rangle}$, is the global fidelity introduced by Bru{\ss} et al\cite{Bruss2}.  Denoting the a priori probability of the state $|{\psi}^{M}_{j}{\rangle}$ by ${\eta}_{j}$, the global fidelity is

\begin{equation}
F_{MN}=\sum_{j=1}^{K}{\eta}_{j}|{\langle}{\psi}_{j}^{N}|{\Phi}_{j}{\rangle}|^{2}.
\end{equation}

Alternative figures of merit have been suggested.  Worth mentioning in particular is the local fidelity, also considered by Bru{\ss} et al, which as its name suggests, is the average fidelity of the individual final states of each of the $N$ subsystems as measured up against the states $|{\psi}_{j}{\rangle}$.  We focus here on the global fidelity since it has an important interpretation in connection with state identification.  Bru{\ss} et al consider the case when the final approximate states $|{\Phi}_{j}{\rangle}$ are generated from the initial states $|{\psi}^{M}_{j}{\rangle}|{\chi}{\rangle}$ by a unitary transformation ${ U}$.  Thus the problem is to vary ${ U}$ such that
\begin{equation}
F_{MN}=\sum_{j=1}^{K}{\eta}_{j}|{\langle}{\psi}_{j}^{N}|{ U}|{\psi}^{M}_{j}{\rangle}|{\chi}{\rangle}|^{2}
\end{equation}
reaches its global maximum.  In the limit as $N{\rightarrow}{\infty}$, the states $|{\psi}_{j}^{N}{\rangle}$ become orthogonal.  In this limit, if we view ${ U}$ as acting to the left in Eq. (2.3), the maximum global fidelity is the maximum probability of discriminating between the states $|{\psi}^{M}_{j}{\rangle}|{\chi}{\rangle}$, or simply $|{\psi}^{M}_{j}{\rangle}$, using an orthogonal measurement.  As a consequence of Kennedy's lemma\cite{Kennedy}, this is actually the maximum discrimination probability for all measurement strategies if, and only if, the $|{\psi}^{M}_{j}{\rangle}$ are linearly-independent.

The analytic solution to this optimisation problem for $K=2$ can be found.  It is convenient here to replace the index $j$ by the simple
binary notation ${\pm}$.  Bru{\ss} et al obtained the maximum global
fidelity for $1{\rightarrow}2$ cloning with equal a priori probabilities.
An important step in their derivation is their observation that the optimum output
states $|{\Phi}_{\pm}{\rangle}$ lie in the subspace spanned by the exact
clones $|{\psi}^{N}_{\pm}{\rangle}$.  Fortunately, this also holds true in the more general case we consider here\cite{Note1}.
The initial state of the entire system is one of $|{\psi}^{M}_{\pm}{\rangle}|{\chi}{\rangle}$, these states having corresponding a priori probabilities ${\eta}_{\pm}$.  A unitary transformation ${ U}$ is applied to generate the states $|{\Phi}_{\pm}{\rangle}={ U}|{\psi}^{M}_{\pm}{\rangle}|{\chi}{\rangle}$.  Let the states $|{\gamma}{\rangle}$ and $|{\delta}{\rangle}$ be an orthonormal basis for the subspace spanned by $|{\psi}^{N}_{\pm}{\rangle}$.  This basis can be chosen in such a way that the exact clones can be expressed as
\begin{equation}
|{\psi}^{N}_{\pm}{\rangle}={\cos}{\theta}|{\gamma}{\rangle}{\pm}{\sin}{\theta}|{\delta}{\rangle}
\end{equation}
where $0{\leq}{\theta}{\leq}{\pi}/4$.  The states to be optimised can be written in the form
\begin{equation}
|{\Phi}_{\pm}{\rangle}={\cos}{\phi}_{\pm}|{\gamma}{\rangle}+{\sin}{\phi}_{\pm}|{\delta}{\rangle}
\end{equation}
for some angles ${\phi}_{\pm}$.  We may, without loss of generality, take ${\phi}_{+}$ and ${\phi}_{-}$ to lie in the first and fourth quadrants respectively.  The aim is to vary these angles such that $F_{MN}$ is maximised, subject to the constraint of unitarity, which implies that ${\phi}_{+}-{\phi}_{-}$ is conserved.  $F_{MN}$ is then given by
\begin{equation}
F_{MN}={\eta}_{+}{\cos}^{2}({\theta}-{\phi}_{+})+{\eta}_{-}{\cos}^{2}({\theta}+{\phi}_{-}).
\end{equation}
Using the method of Lagrange multipliers, it is easy to see that the extrema of $F_{MN}$ with respect to this constraint occur when
\begin{equation}
\frac{{\partial}F_{MN}}{{\partial}{\phi}_{+}}=-\frac{{\partial}F_{MN}}{{\partial}{\phi}_{-}}.
\end{equation}
For this condition to be satisfied, we must have
\begin{equation}
{\tan}2{\theta}=\frac{{\eta}_{+}{\sin}2{\phi}_{+}-{\eta}_{-}{\sin}2{\phi}_{-}}{{\eta}_{+}{\cos}2{\phi}_{+}+{\eta}_{-}{\cos}2{\phi}_{-}}.
\end{equation}
This can be rearranged to give the following expression for ${\cos}^{2}({\phi}_{+}+{\phi}_{-})$:
\begin{equation}
{\cos}^{2}({\phi}_{+}+{\phi}_{-})=\frac{{\cos}^{2}(2{\theta}-{\phi}_{+}+{\phi}_{-})}{1-4{\eta}_{+}{\eta}_{-}{\sin}^{2}(2{\theta}-{\phi}_{+}+{\phi}_{-})}.
\end{equation}

The maximum value of $F_{MN}$ is obtained by substituting the positive root of the optimum ${\cos}^{2}({\phi}_{+}+{\phi}_{-})$, given above, and the root of ${\sin}^{2}({\phi}_{+}+{\phi}_{-})$ with the same sign as ${\eta}_{+}-{\eta}_{-}$ into $F_{MN}$.  Making use of the fact that $F_{MN}$ may be written as
\begin{equation}
F_{MN}=\frac{1}{2}\left\{1+{\cos}({\phi}_{+}+{\phi}_{-}){\cos}(2{\theta}-{\phi}_{+}+{\phi}_{-})+({\eta}_{+}-{\eta}_{-}){\sin}({\phi}_{+}+{\phi}_{-}){\sin}(2{\theta}-{\phi}_{+}+{\phi}_{-})\right\},
\end{equation}
performing the substitution gives the general inequality
\begin{equation}
F_{MN}{\le}\frac{1}{2}\left\{1+\left[1-4{\eta}_{+}{\eta}_{-}{\sin}^{2}(2{\theta}-{\phi}_{+}+{\phi}_{-})\right]^{1/2}\right\}.
\end{equation}

Here we have the least upper bound on the global fidelity for $M{\rightarrow}N$ cloning of two states with arbitrary a priori probabilities.  Evidently it is a decreasing function of ${\eta}_{+}{\eta}_{-}$ and increases as the a priori probabilities differ.  When only one of these is non-zero, the fidelity can attain unity, and perfect cloning is possible.  It reaches its minimum value for equally-probable states.  $F_{MN}$ also decreases with ${\theta}$ and increases ${\phi}_{+}-{\phi}_{-}$, respectively measures of the distinctness of the states $|{\psi}^{N}_{\pm}{\rangle}$ and $|{\psi}^{M}_{\pm}{\rangle}$.

In Section V we will present a quantum logical network which can be used to attain this bound.  For its construction, it will be important to have expressions for the optimum approximate clone states $|{\Phi}_{\pm}{\rangle}$.  As we have already mentioned, these states must be of the form
\begin{equation}
|{\Phi}_{\pm}{\rangle}={\mu}_{\pm}|{\psi}^{N}_{+}{\rangle}+{\nu}_{\pm}|{\psi}^{N}_{-}{\rangle}.
\end{equation}
The coefficients ${\mu}_{\pm}$ and ${\nu}_{\pm}$ are found to be
\begin{eqnarray}
{\mu}_{\pm}&=&\frac{{\sin}({\theta}+{\phi}_{\pm})}{{\sin}2{\theta}}, \\ * {\nu}_{\pm}&=&\frac{{\sin}({\theta}-{\phi}_{\pm})}{{\sin}2{\theta}}.
\end{eqnarray}

In the limit as $N{\rightarrow}{\infty}$, where ${\theta}={\pi}/4$, this inequality becomes
\begin{equation}
F_{M\infty}{\le}\frac{1}{2}\left\{1+(1-4{\eta}_{+}{\eta}_{-}|{\langle}{\psi}^{M}_{+}|{\psi}_{-}^{M}{\rangle}|^{2})^{1/2}\right\}.
\end{equation}
The bound here is the well-known Helstrom bound\cite{Helstrom}, which gives the maximum probability of correctly distinguishing between the two initial states  $|{\psi}_{j}^{M}{\rangle}$.  This is what we would expect on the basis of our discussion of the link between infinite cloning and state identification.  It does, however beg the question: what is the general interpretation of the global fidelity when the number of approximate clones produced is finite?  The natural interpretation is that it is the average probability that a system in the approximate clone state passes a maximal test for being in the exact clone state.  The case of infinite cloning is special since the orthogonality of the exact clone states allows all $K$ of these maximal tests to be carried out simultaneously as a single von Neumann measurement with $K$ outcomes.  

\section{Interpolating Between Approximate and Exact Cloning}
\renewcommand{\theequation}{3.\arabic{equation}}
\setcounter{equation}{0}

In this section we shall be concerned with states having equal a priori probabilities.  Thus, ${\eta}_{+}={\eta}_{-}=0$ and the inequality (2.11) simplifies to
\begin{equation}
F_{MN}{\le}\frac{1}{2}\left\{1+\left[1-{\sin}^{2}(2{\theta}-{\phi}_{+}+{\phi}_{-})\right]^{1/2}\right\}.
\end{equation}
In this special case, ${\phi}_{+}=-{\phi}_{-}$.  We have so far been concerned with deterministic cloning operations.  There does, however, exist the possibility of using probablistic operations, and comparing only some post-selected ensemble with the exact clones.  Duan and Guo\cite{Duanguo1} used this approach to obtain $F=1$, showing that exact clones can be produced with some probability.  Given a system prepared in one of the states $|{\psi}_{\pm}{\rangle}$ and a further system in the neutral state $|{\chi}{\rangle}$, Duan and Guo showed that the maximum probability $P_{12}$ of obtaining the output state $|{\psi}_{\pm}{\rangle}|{\psi}_{\pm}{\rangle}$ is 
\begin{equation}
P_{12}=\frac{1}{1+|{\langle}{\psi}_{+}|{\psi}_{-}{\rangle}|}.
\end{equation}
More generally, given $M$ initial copies in one of the states $|{\psi}_{\pm}{\rangle}$ and an $(N-M)$-particle neutral state, the maximum probability $P_{MN}$ of the $N$ particle output state being $|{\psi}^{N}_{\pm}{\rangle}$, that is, $N$ exact copies of $|{\psi}_{\pm}{\rangle}$, is\cite{Me1}
\begin{equation}
P_{MN}=\frac{1-|{\langle}{\psi}_{+}|{\psi}_{-}{\rangle}|^{M}}{1-|{\langle}{\psi}_{+}|{\psi}_{-}{\rangle}|^{N}}.
\end{equation} 

An interesting question is this: is there a fundamental trade-off between success probability and maximum fidelity?  We can imagine that a cloning device could be constructed which has a higher success rate than an exact cloning machine, though not as high (unity) as the approximate cloning machine of the preceding section.  The fidelity in this case would not reach unity, but for the post-selected ensemble for which the operation succeeds, the fidelity could exceed the bound in Eq. (3.1).  To achieve this, we make use of an inequality, which generalises the bound in (3.3), and relates to an operation we have termed {\em state separation}\cite{Me1}.  As we have shown, given two equiprobable initial states $|{\psi}^{1}_{\pm}{\rangle}$ and two final states $|{\psi}^{2}_{\pm}{\rangle}$ such that 
\begin{equation}
|{\langle}{\psi}^{2}_{+}|{\psi}^{2}_{-}{\rangle}|{\le}|{\langle}{\psi}^{1}_{+}|{\psi}^{1}_{-}{\rangle}|,
\end{equation}
This operation separates the states, since it decreases their overlap.  The maximum attainable value of the probability $P_{S}$ of carrying out the transformation $|{\psi}^{1}_{\pm}{\rangle}{\rightarrow}|{\psi}^{2}_{\pm}{\rangle}$ is given by the bound
\begin{equation}
P_{S}{\le}\frac{1-|{\langle}{\psi}^{1}_{+}|{\psi}^{1}_{-}{\rangle}|
}{1-|{\langle}{\psi}^{2}_{+}|{\psi}^{2}_{-}{\rangle}|}.
\end{equation}
This is a non-unitary operation, or generalised measurement, which involves an interaction with an ancillary system.  We shall discuss its implementation in more detail in sections IV and VI.  An additional point we make here about this state separation bound is that when the final states are orthogonal, we obtain the Ivanovic-Dieks-Peres\cite{Ivanovic,Dieks2,Peres} maximum probability of discriminating {\em without errors} between the states $|{\psi}^{1}_{\pm}{\rangle}$:
\begin{equation}
P_{IDP}=1-|{\langle}{\psi}^{1}_{+}|{\psi}^{1}_{-}{\rangle}|.
\end{equation}
  
To see how state separation may be applied to realise the generalised cloning stategy we have in mind, we observe that the maximum fidelity in (3.1) depends only on two quantities: the modulus of the overlap between the $N$ exact clones, of which it is an increasing function, and that of the initial states, of which it is an increasing function.  It is equal to unity when both overlaps are equal.  Thus, if the initial states are made more distinct to some degree, using state separation whose success probability is bounded by (3.5), the systems post-selected from the successful ensemble can then be transformed unitarily into states having a greater resemblance to the exact clones, as quantified by the global fidelity, than without the initial separation stage.

An inequality relating the success probability and the global fidelity is easily obtained.  Rather than immediately carrying out the approximate cloning transformation upon the initial states $|{\psi}^{M}_{\pm}{\rangle}|{\chi}{\rangle}$, we perform a state separation operation, which transforms these states into some other pair of states, say $|{\tilde \Phi}_{\pm}{\rangle}$.  The overlap between these states is less than that of the initial states, and from the state separation inequality, we obtain
\begin{equation}
|{\langle}{\tilde \Phi}_{+}|{\tilde \Phi}_{-}{\rangle}|{\ge}1-\frac{1-|{\langle}{\psi}^{M}_{+}|{\psi}^{M}_{-}{\rangle}|}{P_{S}}.
\end{equation}
These states may be written as
\begin{equation}
|{\tilde \Phi}_{\pm}{\rangle}={\cos}{\tilde \phi}_{+}|{\gamma}{\rangle}{\pm}{\sin}{\tilde \phi}_{-}|{\delta}{\rangle}.
\end{equation}
The angles ${\tilde \phi}_{\pm}$ are calculated using the results of the optimisation procedure in the preceding section, except that the value of ${\cos}({\tilde \phi}_{+}-{\tilde \phi}_{-})={\langle}{\tilde \Phi}_{+}|{\tilde \Phi}_{-}{\rangle}$ depends on how much we wish to separate the initial states.  Nevertheless, as before, ${\tilde \phi}_{+}-{\tilde \phi}_{-}$ is fixed throughout the optimisation calculation.  The bound on the global fidelity for the post-selected ensemble for which the separation attempt has been successful is given by (3.1) where ${\phi}_{\pm}$ are replaced by ${\tilde \phi}_{\pm}$.  Substitution of (3.7) into this gives the trade-off relation between $F_{MN}$ and $P_{S}$:

\begin{equation}
F_{MN}{\le}\frac{1}{2}\left\{1+|{\langle}{\psi}^{N}_{+}|{\psi}^{N}_{-}{\rangle}|\left(1-\frac{P_{IDP}}{P_{S}}\right)+\frac{1}{P_{S}}\left[(1-|{\langle}{\psi}^{N}_{+}|{\psi}^{N}_{-}{\rangle}|^{2})\left(P_{S}^{2}-(P_{S}-P_{IDP}\right)^{2})\right]^{1/2}\right\}.
\end{equation}
Here, $P_{IDP}$ is given by (3.6), where $|{\psi}^{1}_{\pm}{\rangle}=|{\psi}^{M}_{\pm}{\rangle}$.  

This expression simplifies as $N{\rightarrow}{\infty}$.  Here, as in the discussion of (2.15), the overlap between the exact clones $|{\psi}^{N}_{\pm}{\rangle}$ approaches zero, and we find
\begin{equation}
F_{M{\infty}}{\le}\frac{1}{2}\left\{1+\frac{1}{P_{S}}\sqrt{P_{S}^{2}-(P_{S}-P_{IDP})^{2}}\right\}.
\end{equation}
This inequality, which is derived more directly in \cite{Me1} and \cite{Me2}, relates, as we would expect, to state identification.  Specifically, it corresponds to a state identification measurement with three possible outcomes: inconclusive results, erroneous identification and correct identification of the initial state which is either $|{\psi}^{M}_{\pm}{\rangle}$.  An inconclusive result is found if the separation fails to take place, the probability of which is $1-P_{S}$.  The conditional probabilities that a correct or incorrect result is obtained after successful separation are $F_{M{\infty}}$ and $1-F_{M{\infty}}$ respectively.  This inequality describes a family of measurements, parameterised by $P_{S}$, which interpolates optimally between the Helstrom $(P_{S}=1)$ and Ivanovic-Dieks-Peres $(P_{S}=P_{IDP})$ measurements, and gives a trade-off relation between inconclusive and erroneous results.

\section{Distinguishability Transfer, State Separation and Networks for Exact Cloning}
\renewcommand{\theequation}{4.\arabic{equation}}
\setcounter{equation}{0}

So far we have described two approaches to state-dependent cloning, characterised by distinct figures of merit, and a method of interpolating between their associated limits.  Clearly, it is important to obtain a physical means of carrying out these cloning transformations.  Since all three types of cloning machine involve operations on qubits, a natural approach to the realisation of these cloning machines is to consider them to be quantum computational networks.  Quantum networks for universal cloning have been developed by Bu{\u{z}}ek et al\cite{Buzek2}.  We will consider an $N$-particle quantum register which is prepared in one of the two states $|{\psi}^{M}_{\pm}{\rangle}|{\chi}{\rangle}$.  Our aim is to transform these states according to the specification of each cloning machine, using only basic operations.

Most models of quantum computers involve only three types of operation: unitary transformations on single qubits, unitary interactions between pairs of qubits and measurements on single qubits.  The cloning operations we have described, however, involve collective interactions among all the participating qubits.  For the exact cloning and intermediate cloning machines of the previous section, an ancillary system must also be brought into play in order to realise the probablistic separation transformation.  These collective interactions must be decomposed into single qubit and pairwise operations.  In this section, we show how this may be done for exact cloning, and in the following, we shall construct networks for approximate and the more general hybrid machines we have discussed.  

It is not difficult to render the exact cloning operation in the required form, using a special gate we introduce called the {\em distiguishability transfer gate}.  The motiviation for this comes from the fact that the part of the cloning operation which involves the interaction with an ancilla should be a pairwise interaction.  Therefore, we must transfer all of the information describing which state $|{\psi}^{M}_{\pm}{\rangle}$ the $M$ initial clones are in into one qubit.  This operation must itself be performed using only pairwise and local interactions.
A pairwise interaction gate which, when used repeatedly, results in this desired compression of state information is the distinguishability transfer gate.  We explicitly write the ${\theta}$ dependence of the states $|{\psi}_{\pm}({\theta}){\rangle}={\cos}{\theta}|+{\rangle}+{\sin}{\theta}|-{\rangle}$.  This gate acts as follows:
\begin{equation}
{ D}({\theta}_{1},{\theta}_{2})|{\psi}_{\pm}({\theta}_{1}){\rangle}|{\psi}_{\pm}({\theta}_{2}){\rangle}=|{\psi}_{\pm}({\theta}_{3}){\rangle}|+{\rangle}.
\end{equation}
The operator ${ D}({\theta}_{1},{\theta}_{2})$ is unitary, so we must have
\begin{equation}
{\cos}2{\theta}_{3}={\cos}2{\theta}_{1}{\cos}2{\theta}_{2}.
\end{equation}
This condition, together with $0{\le}{\theta}_{j}{\le}{\pi}/4$, suffices to determine ${\theta}_{3}$ uniquely.  Eq. (4.2) may alternatively be written as 
\begin{equation}
|{\langle}{\psi}_{+}({\theta}_{3})|{\psi}_{-}({\theta}_{3}){\rangle}|=|{\langle}{\psi}_{+}({\theta}_{1})|{\psi}_{-}({\theta}_{1}){\rangle}||{\langle}{\psi}_{+}({\theta}_{2})|{\psi}_{-}({\theta}_{2}){\rangle}|,
\end{equation}
so that the states $|{\psi}_{\pm}({\theta}_{3}){\rangle}$ are seen to be more distinguishable than $|{\psi}_{\pm}({\theta}_{1}){\rangle}$ and $|{\psi}_{\pm}({\theta}_{2}){\rangle}$.  

To obtain an explicit expression for the operator ${ D}({\theta}_{1},{\theta}_{2})$, we must first specify how it transforms states in the subspace orthogonal to that spanned by $|{\psi}_{\pm}({\theta}_{1}){\rangle}|{\psi}_{\pm}({\theta}_{2}){\rangle}$.  A natural completion of the description of ${ D}({\theta}_{1},{\theta}_{2})$  suggests itself if we take the sum and difference of both equations in (4.1), giving
\begin{eqnarray}
{ D}({\theta}_{1},{\theta}_{2})N_{+}({\cos}{\theta}_{1}{\cos}{\theta}_{2}|++{\rangle}+{\sin}{\theta}_{1}{\sin}{\theta}_{2}|--{\rangle})&=&|++{\rangle}, \\ * { D}({\theta}_{1},{\theta}_{2})N_{-}({\cos}{\theta}_{1}{\sin}{\theta}_{2}|+-{\rangle}+{\sin}{\theta}_{1}{\cos}{\theta}_{2}|-+{\rangle})&=&|-+{\rangle},
\end{eqnarray}
where we have called the normalisation factors $N_{\pm}=(2/1{\pm}{\cos}2{\theta}_{3})^{1/2}$.  By choosing states orthogonal to the states on the l.h.s. of (4.4) and (4.5) lying in the subspaces spanned by $|{\pm}{\pm}{\rangle}$ and $|{\pm}{\mp}{\rangle}$ respectively, we obtain
\begin{eqnarray}
{ D}({\theta}_{1},{\theta}_{2})N_{+}({\sin}{\theta}_{1}{\sin}{\theta}_{2}|++{\rangle}-{\cos}{\theta}_{1}{\cos}{\theta}_{2}|--{\rangle})&=&|--{\rangle}, \\ * { D}({\theta}_{1},{\theta}_{2})N_{-}({\sin}{\theta}_{1}{\cos}{\theta}_{2}|+-{\rangle}-{\cos}{\theta}_{1}{\sin}{\theta}_{2}|-+{\rangle})&=&|+-{\rangle}.
\end{eqnarray}
Thus, we have a complete description of the operator ${ D}({\theta}_{1},{\theta}_{2})$, and see that it performs independent rotations on the subspaces spanned by $|{\pm}{\pm}{\rangle}$ and $|{\pm}{\mp}{\rangle}$. In section VI, we show how  may be decomposed into basic operations.  

One remarkable property of the unitary distiguishability transfer operator defined by (4.4)-(4.7) is that it is also Hermitian.  Being its own inverse, this gate can also be used to distribute the distiguishability of the single particle states $|{\psi}_{\pm}({\theta}_{3}){\rangle}$ over the two particle system whose possible states are $|{\psi}_{\pm}({\theta}_{1}){\rangle}|{\psi}_{\pm}({\theta}_{2}){\rangle}$.  In other words, we also have
\begin{equation}
{ D}({\theta}_{1},{\theta}_{2})|{\psi}_{\pm}({\theta}_{3}){\rangle}|+{\rangle}=|{\psi}_{\pm}({\theta}_{1}){\rangle}|{\psi}_{\pm}({\theta}_{2}){\rangle}.
\end{equation}
The distinguishability transfer operator can then either compress or decompress distinguishability. Fig. 1 illustrates the operation of this gate.  As is customary, we have taken time to advance from left to right.  The operator ${ D}$  transfers the distiguishability of the possible states of second particle to those of the first, so that the first particle's final states become more distinct, while the second particle's final states are identical.  A further application of ${ D}$ can perform the reverse process of redistributing the distingushability of the states $|{\psi}_{\pm}({\theta}_{2}){\rangle}$ over both particles.  Both of these properties will turn out to be useful for the implementation of the cloning strategies we have been describing.  Prior to showing this, we briefly remark that, as a consequence of its ability to distribute distinguishability, the transformation ${ D}$ can itself be regarded as a kind of cloning transformation. Taking ${\theta}_{2}={\theta}_{1}$, the operator ${ D}({\theta}_{1},{\theta}_{1})$ acts on a system in one of the states $|{\psi}_{\pm}({\theta}_{3}){\rangle}$, together with a further system in the state $|+{\rangle}$ to produce 2 approximate copies, each in the pure state $|{\psi}_{\pm}({\theta}_{1}){\rangle}$.  Since the final state is a product state, the local fidelity, which we denote by $F_{l}$, and the global fidelity $F$ are simply related.  The local fidelity of the approximate clones obtained this way is 
\begin{equation}
F_{l}={\cos}({\theta}_{3}-{\theta}_{1}),
\end{equation}
irrespective of the a priori probabilities of the states.  The global fidelity $F$ is simply $F_{l}^{2}$.  Although this kind of cloning operation is not optimal, it may still be useful since it leaves the copies uncorrelated, unlike the scheme described in section II.     

\begin{figure}

\epsfxsize=15cm

\epsfysize=3cm

\centerline{\epsffile{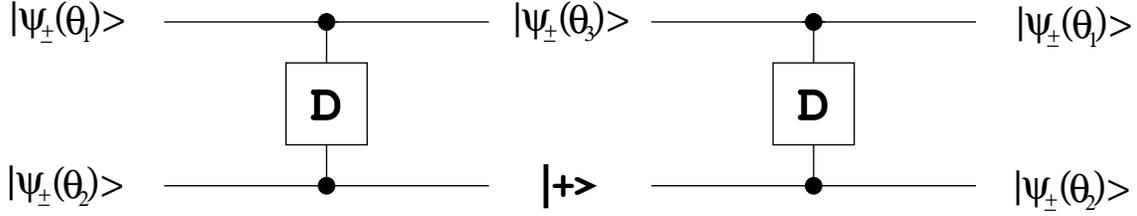}}

\caption{Action of the distinguishability transfer gate.}

\end{figure}  

We now show how the distinguishability transfer operator can be used as a  element in a network for exact $M{\rightarrow}N$ cloning.  The system we consider comprises $N$ qubits, plus one ancilla qubit.  For the sake of simplicity, we take the $(N-M)$ particle neutral state $|{\chi}{\rangle}$ to be $|+{\rangle}|+{\rangle}{\ldots}|+{\rangle}$.   Let us denote the operator which transfers the distinguishability of the possible states of particle $j+1$ to those of particle $j$ by ${ D}_{j}({\theta}_{1},{\theta}_{2})$.  The first stage of the cloning operation then proceeds by application of the operator 
\begin{equation}
{ D}_{1}({\theta}_{1},{\theta}_{M-1}){ D}_{2}({\theta}_{1},{\theta}_{M-2}){\ldots}{ D}_{M-1}({\theta}_{1},{\theta}_{1})|{\psi}^{M}_{\pm}{\rangle}|{\chi}{\rangle}=|{\psi}_{\pm}({\theta}_{M}){\rangle}_{1}|+{\rangle}_{2}|+{\rangle}_{3}{\ldots}|+{\rangle}_{N}.
\end{equation}
Here, we have defined the sequence of angles ${\theta}_{j}$ recursively by
\begin{equation}
{\cos}2{\theta}_{j+1}={\cos}2{\theta}_{1}{\cos}2{\theta}_{j}
\end{equation}
where ${\theta}_{1}={\theta}$ and, as in (4.2), $0{\le}{\theta}_{j}{\le}{\pi}/4$, so that the ${\theta}_{j}$ are uniquely determined by this relation.  The operation (4.10) is simply that which, by pairwise interactions, moves all of the distinguishability from particles $2,{\ldots},M$ to particle 1, as is illustrated in Fig. 2, where $M=3$ and $N=5$.  

The next stage of the cloning operation is to separate the states of the first particle even further, so that the overlap between the final states in (4.10) is equal to ${\langle}{\psi}^{N}_{+}|{\psi}^{N}_{-}{\rangle}$.  This will allow them to be unitarily transformed into exact clones, and it cannot be done with unit probability of success.  The maximum success probability $P_{MN}$ is given by (3.3).  Carrying out this transformation requires the ancilla.  If the ancillary qubit is prepared initally in the state $|+{\rangle}$, the state separation operator ${ S}({\theta}_{M},{\theta}_{N})$ has the following effect:
\begin{equation}
{ S}({\theta}_{M},{\theta}_{N})|+{\rangle}|{\psi}_{\pm}({\theta}_{M}){\rangle}=P_{MN}^{1/2}|+{\rangle}|{\psi}_{\pm}({\theta}_{N}){\rangle}+(1-P_{MN})^{1/2}|-{\rangle}|+{\rangle}.
\end{equation}
The state separation gate is also decomposed into basic elements in section VI.  To complete the description of this operator, again we take the sum and difference of these two equations to obtain:
\begin{eqnarray}
S({\theta}_{M},{\theta}_{N})|++{\rangle}&=&\frac{1}{{\cos}{\theta}_{M}}\left(P_{MN}^{1/2}{\cos}{\theta}_{N}|++{\rangle}+\sqrt{1-P_{MN}}|-+{\rangle}\right), \\ * { S}({\theta}_{M},{\theta}_{N})|+-{\rangle}&=&|+-{\rangle}.
\end{eqnarray}
It is natural to complete the definition of this operator by specifying that it transforms the states $|-{\pm}{\rangle}$ in the following way:
\begin{eqnarray}
S({\theta}_{M},{\theta}_{N})|-+{\rangle}&=&\frac{1}{{\cos}{\theta}_{M}}\left(\sqrt{1-P_{MN}}|++{\rangle}-P_{MN}^{1/2}{\cos}{\theta}_{N}|-+{\rangle}\right), \\ * S({\theta}_{M},{\theta}_{N})|--{\rangle}&=&|--{\rangle}.
\end{eqnarray}
Thus, $S({\theta}_{M},{\theta}_{N})$ so defined is a controlled-rotation operator, either rotating or leaving invariant the state of the ancilla depending on whether the other qubit is in the state $|+{\rangle}$ or $|-{\rangle}$.  Following this interaction, we measure ${\sigma}_{z}$ for the ancilla.  If it is found to be in the state $|+{\rangle}$, then the separation attempt has succeeded.  If, on the other hand, we find that the state is $|-{\rangle}$, the attempt has failed

When the separation attempt succeeds, the two possible final states $|{\psi}_{\pm}({\theta}_{N}){\rangle}_{1}|+{\rangle}_{2}|+{\rangle}_{3}{\ldots}|+{\rangle}_{N}$ have the same overlap as the states of $N$ exact clones, $|{\psi}^{N}_{\pm}{\rangle}$.  Therefore, completion of the cloning procedure can be accomplished by a unitary operation.  This unitary operation must be carried out using pairwise interactions, and here the relevant interaction is the again the distinguishability transfer gate.  Here, we use its ability to decompress the distinguishability of the states of a single qubit, by distributing it over the states of two qubits, as is
 seen in (4.8). 

The actual sequence of operations which completes the cloning procedure is given by the following transformation:
\begin{equation}
{ D}_{N-1}({\theta}_{1},{\theta}_{2}){\ldots}{ D}_{2}({\theta}_{1},{\theta}_{N-2}){ D}_{1}({\theta}_{1},{\theta}_{N-1})|{\psi}_{\pm}({\theta}_{N}){\rangle}_{1}|+{\rangle}_{2}|+{\rangle}_{3}{\ldots}|+{\rangle}_{N}=|{\psi}^{N}_{\pm}{\rangle}.
\end{equation}

As can be seen from Fig. (2), this is simply the application of a sequence of distinguishability transfer gates which distributes the distinguishability of  the single-particle states $|{\psi}_{\pm}({\theta}_{N}){\rangle}$ evenly among the $N$ particles. 

\begin{figure}

\epsfxsize=15cm

\epsfysize=9.4375cm

\centerline{\epsffile{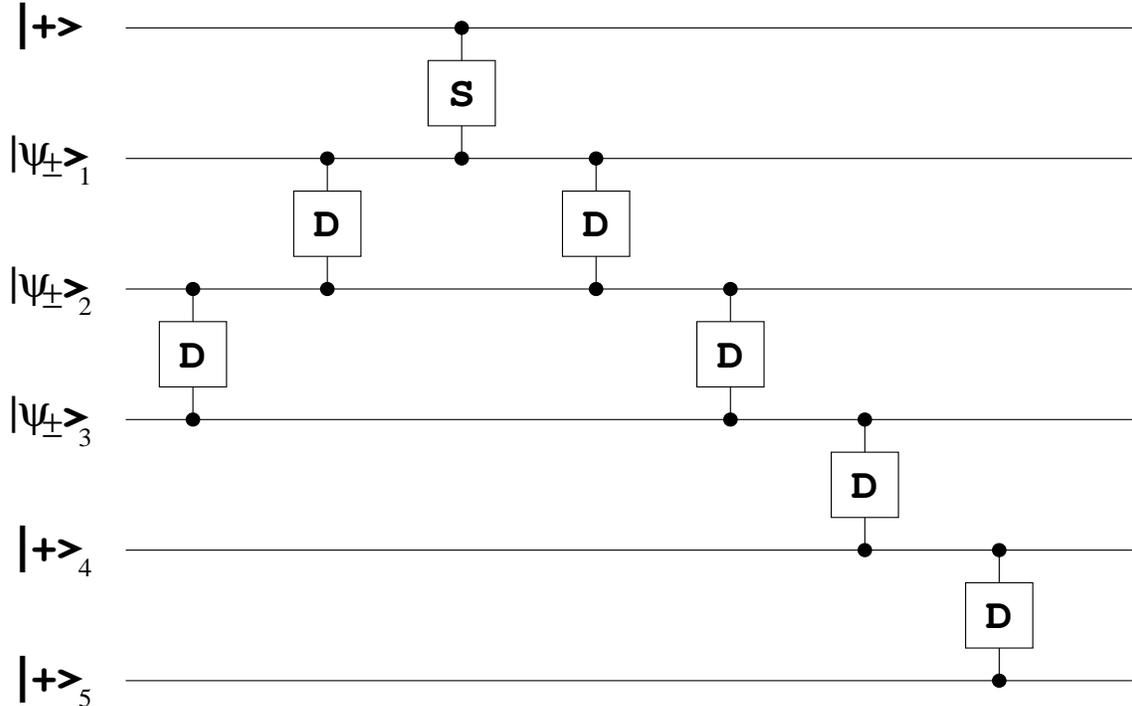}}

\caption{Depiction of a network which may be used for exact cloning.  Here, $M=3$ and $N=5$.  Two distinguishability transfer operators shift the distinguishability of the states of particles 2 and 3 to those particle 1.  An ancilla in the state $|+{\rangle}$ is then used in an attempt to separate the state of particle 1 using the separation gate $S$. If this attempt is successful, distinguishability transfer operators spread the distinguishability of the states of particle 1 evenly over particles 1 to 5.}

\end{figure}

\section{Networks for Optimal Deterministic and Hybrid Cloning}
\renewcommand{\theequation}{5.\arabic{equation}}
\setcounter{equation}{0}

A cloning machine which is designed to maximise the global fidelity by attaining the limit in (2.11) must take as its inputs the states $|{\psi}^{M}_{j}{\rangle}|{\chi}{\rangle}$ and transform them into the states 
\begin{equation}
|{\Phi}_{\pm}{\rangle}={\mu}_{\pm}|{\psi}^{N}_{+}{\rangle}+{\nu}_{\pm}|{\psi}^{N}_{-}{\rangle},
\end{equation}
where ${\mu}_{\pm}$ and ${\nu}_{\pm}$ are given by (2.13) and (2.14).  Since we aim to carry out this transformation using only local operations and pairwise interactions, we again have to transfer all of the information about which of the $M$ particle states $|{\psi}^{M}_{j}{\rangle}$ we have into a single particle.  Therefore, the approximate cloning operation proceeds, as in the exact cloning described above, by repeated pairwise application of the distinguishability transfer operator, resulting in the states shown on the r.h.s. of (4.10).  

Particle 1 is left in one of the states $|{\psi}_{\pm}({\theta}_{M}){\rangle}$.  The next step in the cloning transformation, as is shown in Fig. 3, is a unitary operation on this particle.  There exists a local unitary operator $T$ which performs the transformation
\begin{equation}
{ T}|{\psi}_{\pm}({\theta}_{M}){\rangle}={\mu}_{\pm}|{\psi}_{+}({\theta}_{N}){\rangle}+{\nu}_{\pm}|{\psi}_{-}({\theta}_{N}){\rangle}.
\end{equation}
The unitarity of ${ T}$ can be seen from the fact that the overlaps of the states on the r.h.s. of (5.1) and (5.2) are equal, and from ${\langle}{\psi}_{+}({\theta}_{M})|{\psi}_{-}({\theta}_{M}){\rangle}={\langle}{\Phi}_{+}|{\Phi}_{-}{\rangle}$.

\begin{figure}

\epsfxsize=16cm

\epsfysize=9.4375cm

\centerline{\epsffile{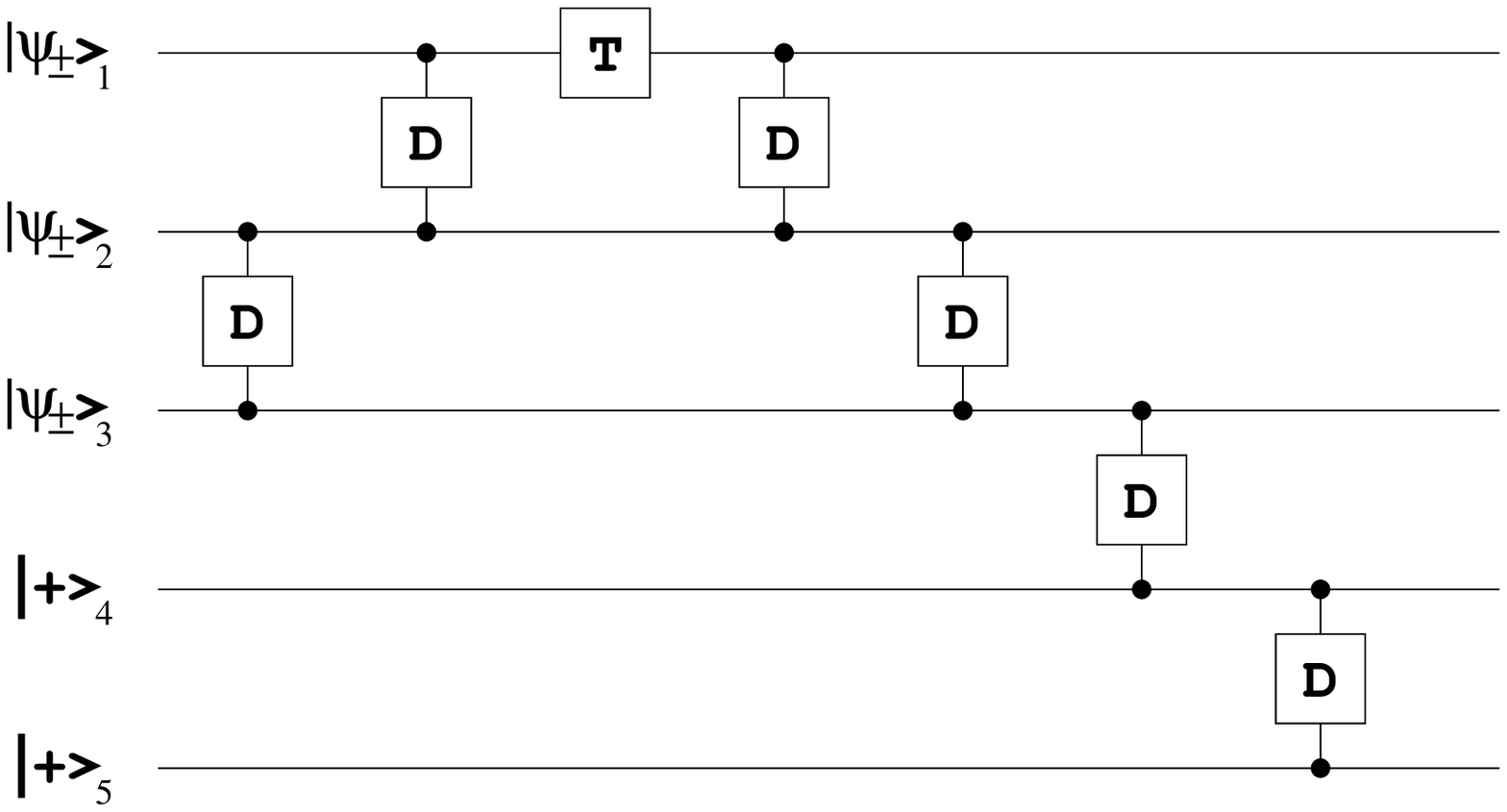}}

\caption{A network for optimum approximate cloning.  The distinguishability of $M$ initial copies of the states is transferred to a single qubit.  A unitary operation $T$ is then performed on this qubit, and application of a further sequence of distinguishability transfer gates results in the optimum multiparticle clone states.}

\end{figure}

The final stage of the cloning operation is simple: it is just the repeated distinguishability decompression used for exact cloning in the previous section.  To see why, we use the linearity of the product of distinguishability transfer operators in (4.17), and apply this transformation to the states $({\mu}_{\pm}|{\psi}_{+}({\theta}_{N}){\rangle}+{\nu}_{\pm}|{\psi}_{-}({\theta}_{N}){\rangle})_{1}|+{\rangle}_{2}|+{\rangle}_{3}{\ldots}|+{\rangle}_{N}$, which results in the desired states $|{\Phi}_{\pm}{\rangle}$ given by (5.1).

Let us now turn to the problem of devising a network which realises the hybrid cloning machine described in section III.  Here, we wish to realise the optimal states $|{\tilde \Phi}_{\pm}{\rangle}$ given by
\begin{equation}
|{\tilde \Phi}_{\pm}{\rangle}={\tilde \mu}_{\pm}|{\psi}^{N}_{+}{\rangle}+{\tilde \nu}_{\pm}|{\psi}^{N}_{-}{\rangle},
\end{equation}
where ${\tilde \mu}_{\pm}$ and ${\tilde \nu}_{\pm}$ are given by (2.13) and (2.14), the same formulas as for ${\mu}_{\pm}$ and ${\nu}_{\pm}$, except that ${\phi}_{\pm}$ are replaced by the angles ${\tilde \phi}_{\pm}$ described in section III. 

The first stage of the cloning operation is again repeated pairwise distinguishability transfer which results in the $N$ particle states $|{\psi}_{\pm}({\theta}_{M}){\rangle}_{1}|+{\rangle}_{2}|+{\rangle}_{3}{\ldots}|+{\rangle}_{N}$.  A pair of intermediate states $|{\Psi}_{\pm}{\rangle}$ are obtained by performing a state separation operation on particle 1.  However, we do not separate the states so much as to produce exact clones.  The effect of the state separation is the transformation 
\begin{equation}
{ S}({\theta}_{M},{\tilde {\theta}})|+{\rangle}|{\psi}_{\pm}({\theta}_{M}){\rangle}=P_{S}^{1/2}|+{\rangle}|{\psi}_{\pm}({\tilde {\theta}}){\rangle}+(1-P_{S})^{1/2}|-{\rangle}|+{\rangle}.
\end{equation}
The angle ${\tilde {\theta}}$ lies in the first quadrant, is defined by ${\cos}2{\tilde {\theta}}=|{\langle}{\Psi}_{+}|{\Psi}_{-}{\rangle}|$ and satisfies ${\theta}_{M}{\le}{\tilde \theta}{\le}{\theta}_{N}$.  If the state separation attempt is successful, the system is left in one of the intermediate states:
\begin{equation}
|{\Psi}_{\pm}{\rangle}=|{\psi}_{\pm}({\tilde {\theta}}){\rangle}_{1}|+{\rangle}_{2}|+{\rangle}_{3}{\ldots}|+{\rangle}_{N}.
\end{equation}
We proceed by analogy with the deterministic approximate cloning, by recognising that there exists a single-qubit unitary operator $T$ such that
\begin{equation}
{ T}|{\psi}_{\pm}({\tilde {\theta}}){\rangle}={\tilde \mu}_{\pm}|{\psi}_{+}({\theta}_{N}){\rangle}+{\tilde \nu}_{\pm}|{\psi}_{-}({\theta}_{N}){\rangle}.
\end{equation}
Thus, the desired states $|{\Phi}_{\pm}{\rangle}$ can be attained by applying the distinguishability decompression operation in (4.17) to the states $|{\Psi}_{\pm}{\rangle}$.  This strategy for hybrid cloning is illustrated in Fig. 4.

\begin{figure}

\epsfxsize=15cm

\epsfysize=9.4375cm

\centerline{\epsffile{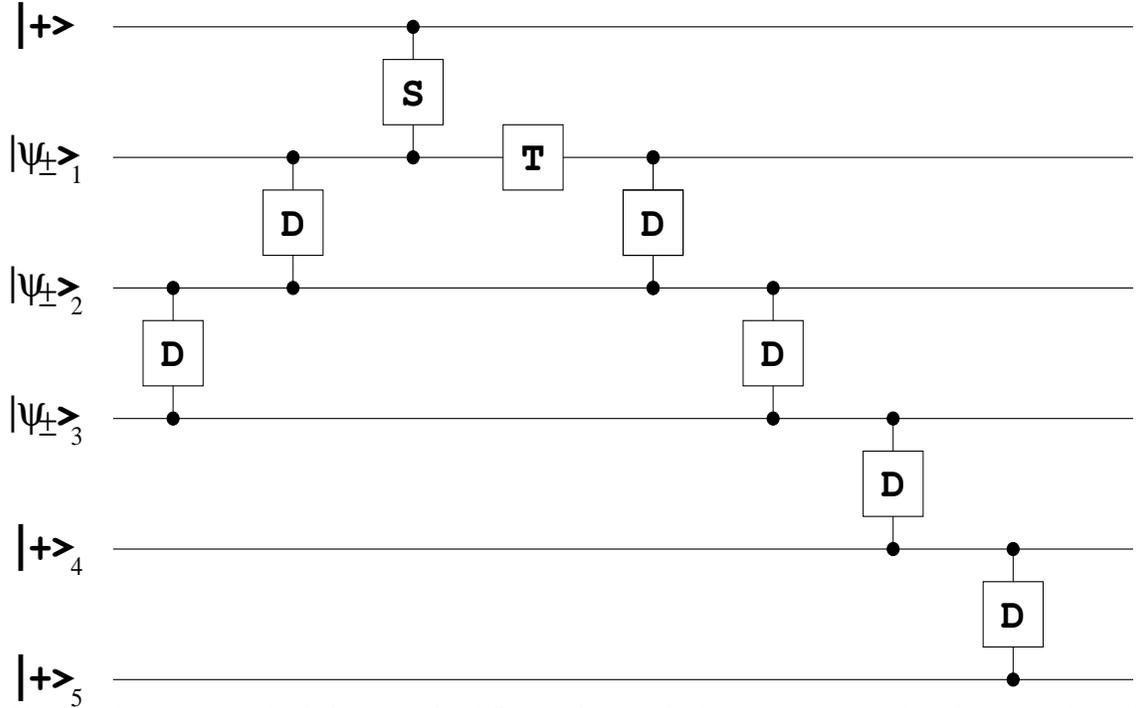}} 

\caption{Network for optimal hybrid cloning.  The difference here with the previous network in Fig. 3 is the inclusion of a state separation gate to the particle to which all distinguishability has been transferred.  With probability bounded by (3.5), this will add further distinguishability to the states of this particle, resulting in improved clones following the final sequence of distinguishability decompression operations.}

\end{figure}

\section{Simulation of the Distinguishability Transfer and State Separation Gates}
\renewcommand{\theequation}{6.\arabic{equation}}
\setcounter{equation}{0}

The previous sections have shown how different quantum cloning strategies may be implemented using only unitary operations and pairwise interactions.  Two gates of the latter kind, the distinguishability transfer and state separation gates, play important roles in the implementation of these schemes.  It is therefore important to determine how these gates may be built. 

It is generally accepted that the fundamental gates of any quantum computer are local unitary (LU) operations on a single qubit, and an interaction gate known the controlled-NOT (CNOT) gate.  Introducing two qubits $a$ and $b$, respectively the control and target qubits, each with the basis $|{\pm}{\rangle}$, the CNOT operator ${ C}_{ab}$ leaves the states $|{-}{\rangle}_{a}|{\pm}{\rangle}_{b}$ invariant, and transforms  $|{\pm}{\rangle}_{b}$ into $|{\mp}{\rangle}_{b}$ when qubit $a$ is in the state $|+{\rangle}_{a}$.  We adopt the schematic convention of Barenco et al \cite{Barenco} by representing the transformations of the target and control qubits by ${\oplus}$ and ${\bullet}$ respectively.

Let us now decompose the distinguishability transfer operator in terms of LU and CNOT gates.  The distinguishability transfer gate ${ D}({\theta}_{1},{\theta}_{2})$, as can be seen from (4.4)-(4.7),  performs independent rotations on the subspaces spanned by $|{\pm}{\pm}{\rangle}$ and $|{\pm}{\mp}{\rangle}$.  These subspaces can be rotated into each other by application of the operator $1{\otimes}{ \sigma}_{x}$.  Let us then denote the operator which performs an orthogonal rotation of the states $|{\pm}{\pm}{\rangle}$ by the angle $\delta$ by ${ Q}({\delta})$.  It acts in the following way:
\begin{equation}
{ Q}({\delta})|{\pm}{\pm}{\rangle}={\pm}{\cos}{\delta}|{\pm}{\pm}{\rangle}+{\sin}{\delta}|{\mp}{\mp}{\rangle}
\end{equation}
and leaves the states $|{\pm}{\mp}{\rangle}$ unchanged.  The operator ${ D}({\theta}_{1},{\theta}_{2})$ can then be written as
\begin{equation}
{ D}({\theta}_{1},{\theta}_{2})={ Q}({\delta}_{1})(1{\otimes}{ \sigma}_{x}){ Q}({\delta}_{2})(1{\otimes}{ \sigma}_{x}),
\end{equation}
where the angles ${\delta}_{i}$ are defined by
\begin{eqnarray}
{\cos}{\delta}_{1}&=&N_{+}{\cos}{\theta}_{1}{\cos}{\theta}_{2}, \;\;\;\;\;\;\;\;{\sin}{\delta}_{1}=N_{+}{\sin}{\theta}_{1}{\sin}{\theta}_{2}, \\ * {\cos}{\delta}_{2}&=&N_{-}{\cos}{\theta}_{1}{\sin}{\theta}_{2}, \;\;\;\;\;\;\;\;{\sin}{\delta}_{2}=N_{-}{\sin}{\theta}_{1}{\cos}{\theta}_{2}.
\end{eqnarray}

The operators ${ Q}({\delta})$ can be expressed in terms of LU and CNOT operations with the aid of the controlled-rotation operators introduced by Barenco et al \cite{Barenco}.  We refer the reader to this reference for a full discussion of this general class of operators.  Here, we restrict our attention to those with 1 control and 1 target qubit.  We also consider only controlled orthogonal rotations rather than general unitary ones.  The type we consider are denoted by ${\wedge}({\delta})$, and act as follows:
\begin{equation}
{\wedge}({\delta})|+{\pm}{\rangle}={\pm}{\cos}{\delta}|+{\pm}{\rangle}+{\sin}{\delta}|+{\mp}{\rangle},
\end{equation}
and leaves the states $|-{\pm}{\rangle}$ unchanged.  If we also define the  unitary Hermitian operator
\begin{equation}
{ E}=(1{\otimes}{ \sigma}_{x}){ C}_{ba}(1{\otimes}{ \sigma}_{x}),
\end{equation}
then one can show that the operators ${\wedge}({\delta})$ and ${ Q}({\delta})$ are related by
\begin{equation}
{ Q}({\delta})=E{\wedge}({\delta})E.
\end{equation}

In \cite{Barenco}, it is shown how the controlled-rotation operator ${\wedge}({\delta})$
can be decomposed in terms of basic operations.  Following the prescription there, we find
\begin{equation}
{\wedge}({\delta}_{i})=(1{\otimes}{ A}_{i}){ C}_{ab}(1{\otimes}{ A}_{i}^{\dagger}),
\end{equation}
where we have introduced the LU operations
\begin{equation}
{ A}_{i}=\frac{1}{\sqrt 2}\left[(1-i{\sigma}_{y}){\cos}\frac{{\delta}_{i}}{2}+(1+i{\sigma}_{y}){\sin}\frac{{\delta}_{i}}{2}\right].
\end{equation}
We now substitute (6.6)-(6.8) into (6.2).  Making use of the identities ${\sigma}_{x}^{2}=1$, $E(1{\otimes}{ \sigma}_{x})E={\sigma}_{x}{\otimes}{\sigma}_{x}$ and of the fact that the CNOT gate is Hermitian, we obtain the following decomposition of $D({\theta}_{1},{\theta}_{2})$:
\begin{equation}
{D}({\theta}_{1},{\theta}_{2})=(1{\otimes}{\sigma}_{x})C_{ba}(1{\otimes}{\sigma}_{x}A_{1})C_{ab}({\sigma}_{x}{\otimes}A^{\dagger}_{1}{\sigma}_{x}A_{2})C_{ab}(1{\otimes}A^{\dagger}_{2}{\sigma}_{x})C_{ba}.
\end{equation}
This sequence of operations is depicted in Fig 5.

\begin{figure}

\epsfxsize=18cm

\epsfysize=3cm

\centerline{\epsffile{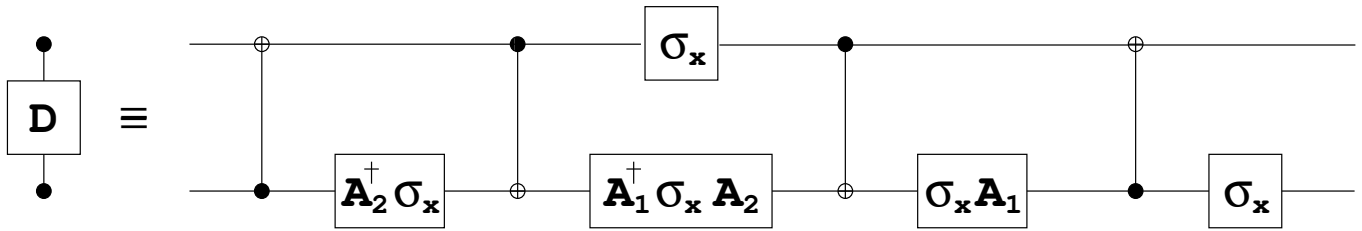}}

\caption{Equivalent network for the distinguishability transfer gate.}

\end{figure}

The state separation gate $S({\theta}_{M},{\theta}_{N})$ is considerably simpler to simulate than the distinguishability transfer gate.  As can be seen from (4.13)-(4.16), this gate is a controlled-rotation gate, where the first and second particles are the target and control qubits respectively.  Let us define the angle ${\gamma}$ by
\begin{equation}
{\cos}{\gamma}=\frac{P_{MN}^{1/2}{\cos}{\theta}_{N}}{{\cos}{\theta}_{M}},\;\;\;\;\;\;\;{\sin}{\gamma}=\frac{\sqrt{1-P_{MN}}}{{\cos}{\theta}_{M}}.
\end{equation}
Then the operator $S$ defined in section IV is simply
\begin{equation}
S({\theta}_{M},{\theta}_{N})=(B{\otimes}1)C_{ba}(B^{\dagger}{\otimes}1),
\end{equation}
where
\begin{equation}
B=\frac{1}{\sqrt 2}\left[(1-i{\sigma}_{y}){\cos}\frac{\gamma}{2}+(1+i{\sigma}_{y}){\sin}\frac{\gamma}{2}\right].
\end{equation}
This decomposition of the state separation gate is shown in Fig. 6.

\begin{figure}

\epsfxsize=9.5cm

\epsfysize=3cm

\centerline{\epsffile{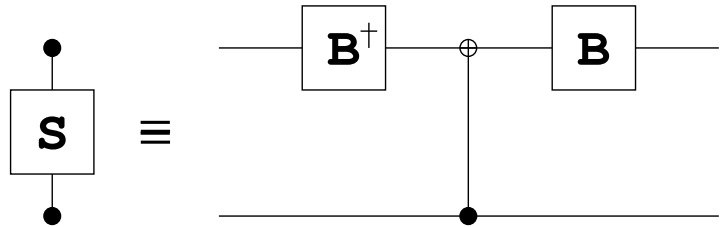}}

\caption{Equivalent network for the state separation gate.}

\end{figure}

\section{Discussion}
In this paper we have obtained some new results pertaining to state-dependent quantum cloning.  In particular, we have generalised the results of Bru{\ss} et al \cite{Bruss2} relating to the optimum global fidelity for two state  $1{\rightarrow}2$ cloning to the case of $M{\rightarrow}N$ cloning and unequal a priori probabilities.  As $N{\rightarrow}{\infty}$, the maximum global fidelity converges to the maximum probability of discriminating between the initial states, which is given by the Helstrom bound.  This connection between the maximum global fidelity for unitary cloning machines and the maximum discimination probability holds generally for more than two states if the states are linearly-independent.  This raises the question of whether or not the global fidelity is the most appropriate figure of merit if the states are linearly-dependent.

It is also interesting to note that {\em exact} cloning and {\em unambiguous} discrimination of states, the latter being equivalent to {\em infinite} exact cloning\cite{Duanguo2,Me1,Me3}, are also only possible for linearly-independent states.  These states then appear to have a special status with regard to state-dependent cloning in general. 

We have also constructed a continuous family of cloning strategies, which optimally interpolates between deterministic, approximate cloning and probablistic exact cloning.  During the cloning procedure, an attempt is made to separate the states.  The less the degree of separation, the higher the probability of success.  Better clones can be obtained from separated states, and it is this probability, or equivalently, the degree of separation which parameterises this family of strategies.

The implementation of the cloning strategies we have described was discussed for the remainder of the paper.  We showed how they may be implemented using local unitary operations and pairwise interactions.  The networks we have described are made conceptually simple and scalable by two special gates: the state separation and distinguishability transfer gates.  We have also have shown how these gates can be decomposed into basic operations. 

\section*{Acknowledgements}
We thank Masahide Sasaki and Richard Josza for helpful discussions.  This work was supported by the UK Engineering and Physical Sciences Research Council (EPSRC).

\end{document}